%% file: main.tex
\documentclass[conference,11pt]{IEEEtran}

\makeatletter
\def\footnoterule{\relax%
  \kern-5pt
  \hbox to \columnwidth{\hfill\vrule width 0.8\columnwidth height 0.4pt\hfill}
  \kern4.6pt}
\makeatother

\usepackage{multirow}
\usepackage[utf8]{inputenc}
\usepackage{lipsum}
\usepackage{multicol}
\usepackage{graphicx}

\usepackage{listings}
\usepackage{xcolor}
\usepackage{hyperref}
\let\svthefootnote\thefootnote
\newcommand\freefootnote[1]{%
  \let\thefootnote\relax%
  \footnotetext{#1}%
  \let\thefootnote\svthefootnote%
}
\title{What Does the Post-Moore Era Mean for Research Software Engineering?}
\author{
\IEEEauthorblockN{Kazutomo Yoshii}
\IEEEauthorblockA{Mathematics and Computer Science\\
Argonne National Laboratory\\
Lemont, IL 60439}
}

\thanks{thanks}

\usepackage{graphicx}

\begin{document}
\maketitle

\input{narrative}


\section*{Acknowledgments}

This work is based on work supported by the U.S. Department of Energy, Office of Science, under contract  DE-AC02-06CH11357.

\clearpage

\bibliographystyle{ieeetr}
\bibliography{refs}

\end{document}

%% file: narrative.tex
\section*{Abstract}

We are entering the post-Moore era where we no longer enjoy the free ride of the performance growth from simply shrinking the transistor features. However, this does not necessarily mean that we are entering a dark era of computing. On the contrary, sustaining the performance growth of computing in the post-Moore era itself is cutting-edge research. Concretely, heterogeneity and hardware specialization are becoming promising approaches in hardware designs. However, these are paradigm shifts in computer architecture. So what does the post-Moore era mean for research software engineering? This position paper addresses such a question by summarizing possible challenges and opportunities for research software engineering in the post-Moore era. We then briefly discuss what is missing and how we prepare to tackle such challenges and exploit opportunities for the future of computing.


\section*{Heterogenity and HPC}

First of all, we must compensate for the increasingly manifested inefficiencies on general-purpose processors~\cite{hameed2010understanding} in the post-Moore era, with innovations in algorithms, software techniques, and computer architectures. At the cusp of the post-Moore era, we are now observing higher-level heterogeneity in the node architecture design in pre-exascale or exascale systems. Unfortunately, those new heterogeneous architectures will not magically accelerate existing high-performance computing (HPC) applications unless significantly modified. Such modification is to applications, libraries, compilers, and system software. Software stack needs to be changed drastically to utilize the massive parallelism offered by the heterogeneous architecture with general-purpose computation on graphics processing units. It is no exaggeration to say that the success of exascale depends on software engineering in the realistic view. Numerous efforts are being made both in academia and industry to accommodate the paradigm shift in computer architecture. We believe that research software engineering plays a critical role in exascale.

\section*{Hardware specialization and scientific instruments}

Besides heterogeneity, such as in exascale, hardware specialization or custom hardware development is another promising direction in the post-Moore era to improve the performance per transistors and enable the integration of HPC and scientific instruments.
AI accelerators are an excellent example of hardware specialization, which significantly improves the performance per transistor for AI workloads.  Hardware specialization is also critical for tight integration between HPC and scientific instruments for future scientific laboratories. For example, ever-increasingly data volume from scientific instruments would overwhelm the network or make it impossible to send out all data to data analysis systems. Theoretically, we must process data as close as sensors or acquisition systems in a data flow manner while offloading to general-purpose processors adds prohibitive overheads for data processing. We could develop on-chip processing capability for detector application-specific integrated circuit (ASIC) as an ideal approach, making the data path shortest, or develop processing capability in field-programmable gate arrays (FPGA) directly connected to sensors. However, these approaches require actual custom hardware development at the digital circuit level, outside of average research software engineers' expertise. 

\section*{Hardware development ecosystems}

Counterintuitively, all development processes in hardware development are done via software, although the outcomes are physical devices (e.g., computer chips). We describe digital circuits using hardware description languages (HDL) such as Verilog and translate HDL codes into an integrated-circuit layout database file such as GDS-II for fabrication, using EDA tools, hardware compilers.
Historically, a few companies dominate hardware design ecosystems for ASICs, and the licensing fee for their commercial tools can often be a significant barrier to entry for many organizations.
Many hardware-related software tools may be outdated or only available as closed commercial software. Furthermore, it requires significant efforts to master each tool and hardware library, but the real problem is that there is no guarantee that acquired knowledge is transferable to other tools. Interestingly, for the last couple of years, the landscape of hardware development ecosystems (hardware languages, electronic design automation (EDA) tools, process design kits) are rapidly changing, partially due to strong tailwinds created by open-source hardware ecosystems such as RISC-V instruction set~\cite{asanovic2014instruction}, emerging hardware description languages~\cite{truong2019golden}, OpenRoad EDA tool~\cite{kahng2020open},  Google-Skywater process design kits. 
The recent open-source hardware movement could encourage software engineers to be involved in hardware development.  Additionally, opportunities for us are not only from the benefit of custom hardware but also in hardware development ecosystems themselves (e.g., parallelizing hardware compilers and simulators, open-source hardware library development).

\section*{The needs of hardware-minded research software engineers}

Admittedly, funding can shape our direction and expertise. Unfortunately, until recently, there is little direct funding to support hardware and architecture research. Of course, this can explain that our hardware expertise is weak. However, in the last couple of years, the number of funding opportunities related to software/hardware co-design, new architecture development, and post-Moore research, in general, seems to be increasing gradually, which can be an excellent tailwind for our future direction. Even though we are into custom hardware development, 
hardware/software co-design is crucial for our future computing needs in HPC. Since we design new architectures, collaborating with computer architects and hardware engineers in industries, the co-design activity requires profound knowledge and insight into computer architectures 
However, unfortunately, hardware-minded research software engineers are a minority in the HPC community at this point. We are strongly concerned about this fact.


\section*{How can we fill the needs?}

It is challenging to hiring hardware experts simply because we, DOE research laboratories, are hardly their ideal workplace.
Moreover, hiring hardware experts may not solve this problem because hardware development ecosystems are rapidly changing, as previously mentioned. We are interested in a relatively small-scale, rapid custom hardware development rather than large-scale conventional hardware development in large companies such as Intel. Thus, we need software-minded hardware experts or hardware-minded software engineers. Besides, emerging hardware description languages are based on modern software languages and practice. For example, chisel hardware description language is based on Scala, fully open-source software that offers sufficient features for efficiently describing algorithmically complex hardware designs. Moreover, chisel has brought modern software practices to hardware development. In that sense, training software experts to develop hardware circuits could be faster than training hardware experts to learn modern software paradigms.

While compiling, running, testing, and debugging user codes are relatively straightforward in software, comparable actions in hardware development are far more complex than software. Additionally, hardware development ecosystems can be challenging to set up and use. While open-source hardware development ecosystems address such problems, a shared development environment is critical for accelerating training and collaboration. To this end, we have been utilizing Chameleon cloud~\cite{keahey2020lessons} as a shared environment for mentoring students and conducting hardware-related research collaboration  (e.g., thermal-aware/power-aware computing, reconfigurable computing). In addition, we are currently setting up an open-source ASIC-development EDA tool experimentally in the Chameleon cloud.

Like other disciplines, building topic-focused communities is essential for discussing technical challenges, shared testbeds, and funding opportunities.  In addition, we have been running bird-of-feather sessions, panel sessions, workshops in major conferences (e.g., supercomputing conference) on field-programmable gate arrays or FPGAs for HPC. However, as of this writing, no noticeable community for hardware specialization for HPC or integration with scientific instruments exists.


%% file: main.bbl
\begin{thebibliography}{1}

\bibitem{hameed2010understanding}
R.~Hameed, W.~Qadeer, M.~Wachs, O.~Azizi, A.~Solomatnikov, B.~C. Lee,
  S.~Richardson, C.~Kozyrakis, and M.~Horowitz, ``Understanding sources of
  inefficiency in general-purpose chips,'' in {\em Proceedings of the 37th
  annual international symposium on Computer architecture}, pp.~37--47, 2010.

\bibitem{asanovic2014instruction}
K.~Asanovi{\'c} and D.~A. Patterson, ``Instruction sets should be free: The
  case for {RISC-V},'' {\em EECS Department, University of California,
  Berkeley, Tech. Rep. UCB/EECS-2014-146}, 2014.

\bibitem{truong2019golden}
L.~Truong and P.~Hanrahan, ``A golden age of hardware description languages:
  applying programming language techniques to improve design productivity,'' in
  {\em 3rd Summit on Advances in Programming Languages (SNAPL 2019)}, Schloss
  Dagstuhl-Leibniz-Zentrum fuer Informatik, 2019.

\bibitem{kahng2020open}
A.~B. Kahng, ``Open-source {EDA}: If we build it who will come?,'' in {\em
  Proc. 28th IFIP/IEEE International Conference on Very Large Scale Integration
  (VLSI-SoC)}, 2020.

\bibitem{keahey2020lessons}
K.~Keahey, J.~Anderson, Z.~Zhen, P.~Riteau, P.~Ruth, D.~Stanzione, M.~Cevik,
  J.~Colleran, H.~S. Gunawi, C.~Hammock, J.~Mambretti, A.~Barnes, F.~Halbach,
  A.~Rocha, and J.~Stubbs, ``Lessons learned from the chameleon testbed,'' in
  {\em Proceedings of the 2020 USENIX Annual Technical Conference (USENIX ATC
  '20)}, USENIX Association, July 2020.

\end{thebibliography}
